\documentclass[twocolumn]{jpsj3} %% two-column layout
\title{Polaronic behavior of photoelectron spectra of Fe$_3$O$_4$ revealed by both hard X-ray and extremely low energy photons}

\author{Masato \textsc{Kimura}$^{1}$, Hidenori \textsc{Fujiwara}$^{1}$, Akira \textsc{Sekiyama}$^{1, 2}$, Junichi \textsc{Yamaguchi}$^{1}$, Kazumasa \textsc{Kishimoto}$^{1}$, Hiroshi \textsc{Sugiyama}$^{1}$, 
Gen \textsc{Funabashi}$^{1}$, Shin \textsc{Imada}$^{3}$,\\ Satoshi \textsc{Iguchi}$^{4}$, Yoshinori \textsc{Tokura}$^{4}$, 
Atsushi \textsc{Higashiya}$^{2, 5}$, Makina \textsc{Yabashi}$^{2, 6}$,\\ Kenji \textsc{Tamasaku}$^{2}$, Tetsuya \textsc{Ishikawa}$^{2}$, Takahiro \textsc{Ito}$^{7}$\thanks{Present address: Graduate School of Engineering, Nagoya University, Nagoya, Aichi, 464-8603, Japan}, Shin-ichi \textsc{Kimura}$^{7}$,\\ and Shigemasa \textsc{Suga}$^{1, 2}$}

\inst{$^{1}$Division of Materials Physics, Graduate School of Engineering Science, Osaka University,\\ Toyonaka, Osaka 560-8531, Japan. \\
$^{2}$SPring-8/Riken, Sayo, Hyogo 679-5148, Japan \\
$^{3}$Department of Physical Sciences, Ritsumeikan University, Kusatsu, Shiga 525-8577, Japan \\
$^{4}$Department of Applied Physics, University of Tokyo, Bunkyoku, Tokyo 113-8656, Japan \\
$^{5}$Industrial Technology Center of Wakayama Prefecture, Wakayama 649-6261, Japan \\
$^{6}$SPring-8/JASRI, Sayo, Hyogo 679-5198, Japan \\
$^{7}$UVSOR Facility, Institute of Molecular Science, Okazaki, Aichi, 444-8585, Japan }

\abst{Hard X-ray and extremely low energy bulk-sensitive photoelectron spectroscopy has been performed in the temperature range of 100--330 K for Fe$_3$O$_4$. In the high temperature phase just above the Verwey transition, the intensity at the Fermi level ($E_F$) is still negligible, but it increases gradually with further increasing the temperature (250 K, 330 K) in consistence with the temperature dependence of the conductivity. The spectral behaviors near E$_F$ with temperature are well explained by the model, which takes the polaron effect into account. }

\kword{strongly correlated electron system, Verwey transition, Magnetite,\\ photoelectron spectroscopy, polaron}

\begin{document}
\maketitle
Fe$_3$O$_4$ (magnetite) is known as one of the oldest magnetic oxides with high potential for applications to spin-electronics (spintronics). Many theoretical results have predicted the half metallic ferromagnetism (HMF) in Fe$_3$O$_4$ with a conductivity by minority spin electrons and the semiconducting behavior by the majority spin channel.\cite{lda1,lda2,lda3,lda4} Recently, the experimental evidence for the HMF is given for the epitaxial Fe$_3$O$_4$ (111) thin films at room temperature (RT) by means of the spin-resolved photoelectron spectroscopy in the photon energy ($h\nu $) range of 20--60 eV.\cite{spes1,spes2}

Fe$_3$O$_4$ is well known to show the Verwey transition across $T_V\sim $ 123 K,\cite{verwey1} where the first order transition of conductivity takes place with its decrease by about two orders of magnitude accompanied with the lattice distortion.\cite{conductivity1} There are controversial discussions on the origin of this transition.\cite{verwey2,MIT,con-cal,discuss1,discuss2,discuss3} For example, Fe$_3$O$_4$ is thought to be metallic above $T_V$, although  the temperature dependence of the conductivity in Fe$_3$O$_4$ suggests the insulating character below $\sim $300 K. The photoelectron spectral weight near the Fermi level ($E_F$) has been found to be very weak even above $T_V$\cite{UPS1,UPS2,SXPES} in contrast to the results of the band calculation. The  insulator to insulator transition across $T_V$ with increasing the temperature is proposed by conventional photoelectron spectroscopy (PES) performed in the $h\nu $ range of 20--100 eV.\cite{UPS1} On the other hand, the insulator to metal transition (IMT) is proposed by a high resolution photoemission study by Chainani $et$ $al.$\cite{UPS2} Very recently more bulk sensitive soft X-ray PES (SXPES) has been performed at $h\nu $ = 707.6 eV under the Fe 2p--3d resonance excitation, and  revealed the insulator to insulator transition across $T_V$.\cite{SXPES}

As well known nowadays, the surface electronic structures are much different from the bulk electronic structures\cite{CePES1,CePES2,YbAl3} in strongly correlated electron systems because $U/t$ ($U$ : electron correlation energy, $t$ : intersite hopping energy) is much larger in the surface region compared with that in the bulk region. Even in the SXPES, however, the surface spectral weight near $E_F$ can be as high as 20--40 $\%$. Therefore more bulk-sensitive PES studies are desired for a long time. In this decade, hard X-ray PES (HAXPES) has widely been recognized as a highly bulk-sensitive technique because of the long inelastic mean-free path (IMFP) and intensively applied to various strongly correlated electron systems.\cite{HAXPES1,HAXPES2,HAXPES3} The IMFP reaches up to $\sim $100 \AA\  at $h\nu\sim $ 8 keV.\cite{IMFP} Meanwhile extremely low energy PES (ELEPES) with a laser excitation near 7 eV turned out to provide rather bulk sensitive results.\cite{ELEPES1,ELEPES2} The IMFP for ELEPES is expected to reach values comparable to that for HAXPES under certain conditions. We have here employed both HAXPES and ELEPES for studying the electronic structures of Fe$_3$O$_4$ across the Verwey transition and in high temperatures up to 330K, where we confirmed that the behavior of bulk PES can be consistently understood by considering the polaron effects.

The experiments were performed on single crystalline Fe$_3$O$_4$ grown by the floating-zone method. HAXPES measurements were performed at BL19LXU in SPring-8\cite{SP8_19} with an MB Scientific (MBS) A1-HE hemispherical analyzer. The ELEPES measurements were performed by use of synchrotron radiation at BL7U in UVSOR\cite{UVSOR_7} with an MBS-A1 analyzer. The photon energy employed in HAXPES (ELEPES) was 7941 eV (7.5 eV) and the energy resolution was set to $\sim $120 meV ($<$10 meV). Clean surfaces were obtained by fracturing in situ at 140 K (room temperature) in HAXPES (ELEPES). The Fermi level was calibrated by the Fermi edge of Au electrically connected to the sample. The HAXPES (ELEPES) measurements were performed at 100, 140, 250, and 330 K (140, 250, and 330 K). 

\begin{figure}[t]
\begin{center}
\includegraphics[width=7cm]{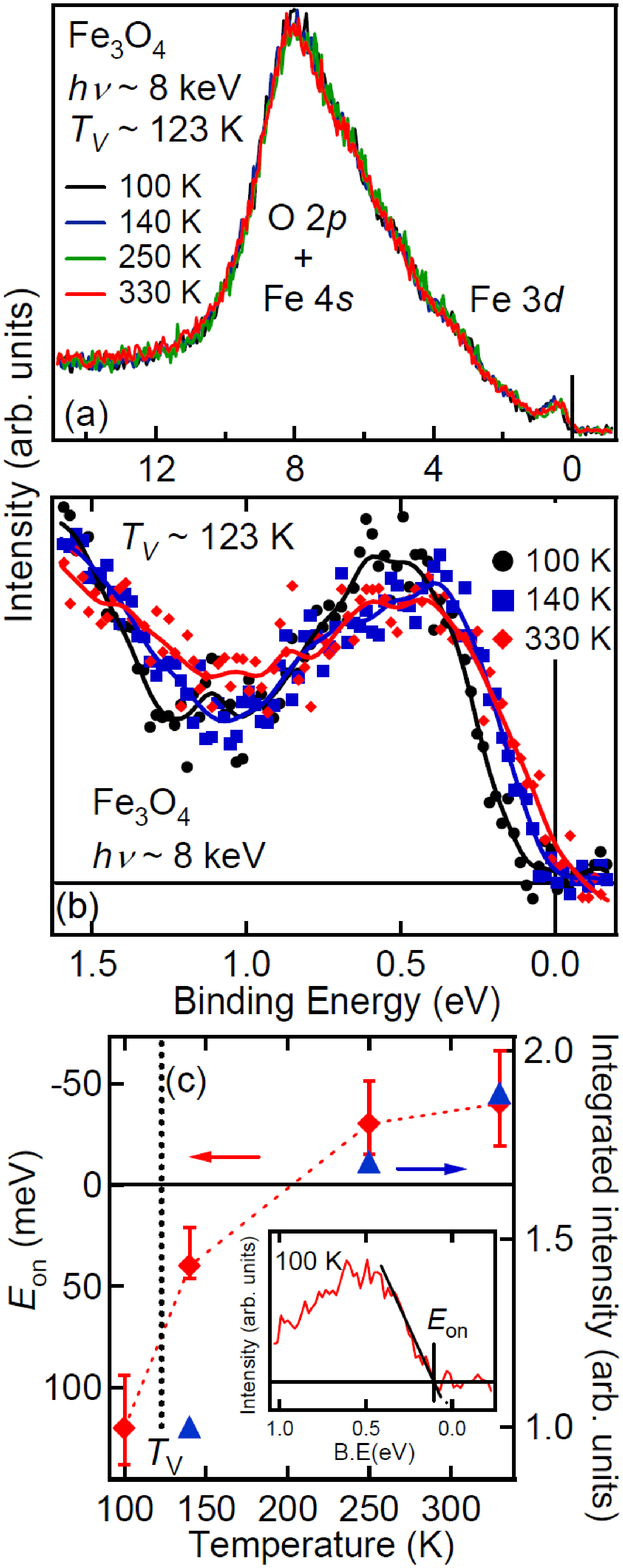}
\end{center}
\caption{(color online) (a) Temperature dependence of the hard X-ray photoemission (HAXPES) spectra of the valence band of Fe$_3$O$_4$ in the wide $E_B$ region normalized by the area above $E_B$ = 3 eV. (b) Temperature dependence of the HAXPES valence band spectra of Fe$_3$O$_4$ near $E_F$ measured with the total resolution of 120 meV. The dots and solid lines show the experimental and the smoothed data, respectively. (c) Temperature dependence of the energy ($E_{on}$) of the spectral onset (red diamonds) (see inset for definition) and the spectral intensity at $E_F$ (blue triangles) above $T_V$ tentatively obtained by integrating the photoelectron intensity from -0.1 to 0.1 eV.}
\label{f1}
\end{figure}

Figure 1(a) displays the temperature dependence of the HAXPES valence-band spectra of Fe$_3$O$_4$ in both low temperature (LT) phase (100K) and high temperature (HT) phase (140, 250, and 330 K) with respect to $T_V$ measured with the resolution of $\sim $200 meV. The broad band ranging from 5 to 12 eV is ascribed to the so-called O 2p valence band. One notices that the intensity near 8 eV is much stronger than the intensity near 4 eV in strong contrast to the results of low energy and soft X-ray PES.\cite{UPS1,UPS2,SXPES} According to the photoionization cross section\cite{cross_section} the strong intensity near 8 eV is ascribed to the Fe 4s component hybridized with the O 2p component similar to vanadium oxides,\cite{VO2,V6O13} whereas the structures in the binding energy ($E_B$) below 5 eV are either Fe 3d--O 2p anti-bonding state or the O 2p non-bonding state. It should be noticed that the transition metal 4s + O 2p components in 5--12 eV in the HAXPES spectra have less temperature dependence than those for most vanadium oxides with metal-to-insulator transitions.\cite{VO2,V6O13} As shown in Fig. 1(b), however, a clear shift of the spectral weight is observed near the Fermi level ($E_F$) across the Verwey transition in the spectra measured with higher resolution of 120 meV. In the LT phase (100 K), the spectrum has the small energy gap as judged from the position of the low-$E_B$ threshold and negligible intensity at $E_F$, as expected for an insulator. In the HT phase just above $T_V$ (at 140 K), the spectral onset becomes shifted towards $E_F$, but the intensity just at $E_F$ is still very weak as already seen in SXPES.\cite{SXPES} Furthermore, spectral change was observed with further increasing temperature up to 330 K in HT phase. The integrated spectral intensity between -0.1 and 0.1 eV is plotted as a function of temperature in Fig. 1(c) by the blue triangle marks. It increases gradually with temperature in HT phase and the enhancement of the spectral weight between 250 and 330 K is smaller than that between 140 K and 250 K. For a more quantitative discussion, we phenomenologically define the spectral onset $E_{on}$ as the intersection of the leading edge with the zero intensity base line as shown in the inset of Fig. 1(c). $E_{on}$ is located below $E_F$ at 100 K in the LT phase and just above $T_V$ (at 140K) as expected for an insulator, whereas it is above $E_F$ at 250 K and 330 K. 

\begin{figure}
\begin{center}
\includegraphics[width=7cm]{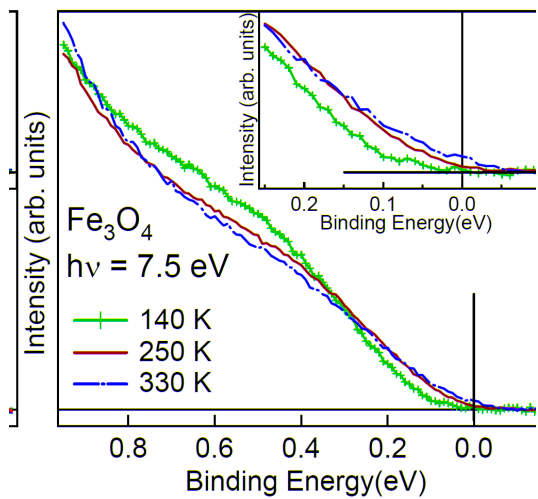}
\end{center}
\caption{(color online) Temperature dependence of the ELEPES spectra for Fe$_3$O$_4$ in the high-temperature phase. The inset shows the expanded ELEPES spectra near $E_F$.}
\label{f2}
\end{figure}

In order to see the detailed temperature dependence of the intensity at $E_F$ in the HT phase, ELEPES has been performed in the temperature range of 140-330 K for Fe$_3$O$_4$ with much better energy resolution ($<$10 meV). Figure 2 displays the temperature dependence of the ELEPES spectra near $E_F$ of Fe$_3$O$_4$ in the HT phase. The spectra are normalized by the area with $E_B$ larger than 1.0 eV since this region dominated by the O 2p states had no temperature dependence in HAXPES spectra. The shapes of ELEPES spectra are much different from those of HAXPES spectra. For example, the intensity increases still above 0.8 eV in contrast to the HAXPES. According to the photoionization cross section, the ratio between the O 2p and Fe 3d components is much larger in ELEPES spectra than in the HAXPES.\cite{cross_section} Therefore, the spectral shapes of ELEPES mainly reflect the O 2p component hybridized with the Fe 3d states. At 140 K, the intensity is very weak at $E_F$, but it becomes finite and increases with temperature (inset of Fig. 2) in consistence with the HAXPES result. The temperature dependence of the intensity at $E_F$ can not be simply explained by the Femi-Dirac distribution function. Moreover, the intensity at around 0.6 eV is gradually suppressed with increasing temperature. These temperature dependence clearly indicates that the electronic states of Fe$_3$O$_4$ change even within the HT phase. These spectral changes can well explain the unusual thermal behavior of the conductivity above $T_V$, which increases toward 330 K. One notices here that the sharp Fermi edge is never observed in Fe$_3$O$_4$ at 330K. In the LDA calculation,\cite{lda3} however, large density of states (DOS) is predicted near $E_F$ due to the Fe 3d (hybridized with O 2p) down-spin states in contrast to the experimental results. 

This discrepancy between the PES spectra and LDA calculation results can be explained by the formation of the polaronic quasiparticle due to the electron-phonon coupling, which is often discussed in HT phase of Fe$_3$O$_4$. Degiorgi $et$ $al.$ explained the unusual temperature behavior of the conductivity above $T_V$ by using the small polaron model which also takes the polaronic short range order into account.\cite{pol.model1} The mid-infrared polaron peak predicted from a polaronic binding energy of $\varepsilon _p$ $\sim $ 300 meV has really been observed in the optical conductivity for Fe$_3$O$_4$.\cite{OC1,OC2} Strong coupling to the lattice vibration and the formation of small polarons have also been inferred from the small magnitude of carrier mobility.\cite{carrier_mobility} Besides, the temperature dependence of the HAXPES and ELEPES spectra near $E_F$ of Fe$_3$O$_4$ is very similar to the spectral changes of (TaSe$_4$)$_2$I and K$_{0.3}$MoO$_3$, which are known as the quasi-one-dimensional Peierls systems with polaronic effects.\cite{peierls_system1,peierls_system2} The previous SXPES measurements have also suggested the formation of the small polaron in HT phase of Fe$_3$O$_4$ by using of the theoretical polaronic model of Alexandrov and Ranninger,\cite{pol.model2} in which the electron removal spectrum of a system with strong electron phonon coupling can be written as
\begin{equation}
I(E_B)\propto e^{-g^2}\tilde {N}_p(E_B)+\sum_{n=1}^\infty e^{-g^2}\frac{g^{2n}}{n!}\tilde {N}_p(E_B-n\omega _0).
\end{equation}
The first term describes the polaronic quasiparticle band ($\tilde {N}_p(E_B)$) renormalized by a factor $e^{-g^2}$, where $g^2=\varepsilon _p/\omega _0$ is a dimensionless electron-phonon coupling strength given by $\varepsilon _p$ (polaronic binding energy) divided by $\omega _0$ (characteristic phonon energy). The spectral intensity is partially transferred to multiple phonon side bands shifted by $n\omega _0$ toward higher binding energies. If the coupling constant $g^2$ becomes stronger, the polaron quasiparticle weight is exponentially suppressed in the photoemission spectrum. This polaronic model could fit the experimental spectra of SXPES at 180 K with parameters of $g^2$ $\sim $ 5 and $\omega _0$ $\sim $ 70 meV in consistence with both the polaronic binding energy ($\sim $300 meV) and the effective mass ($\sim $100--200$m_0$) derived from the infrared spectroscopy.\cite{SXPES}

\begin{figure}[t]
\begin{center}
\includegraphics[width=7cm]{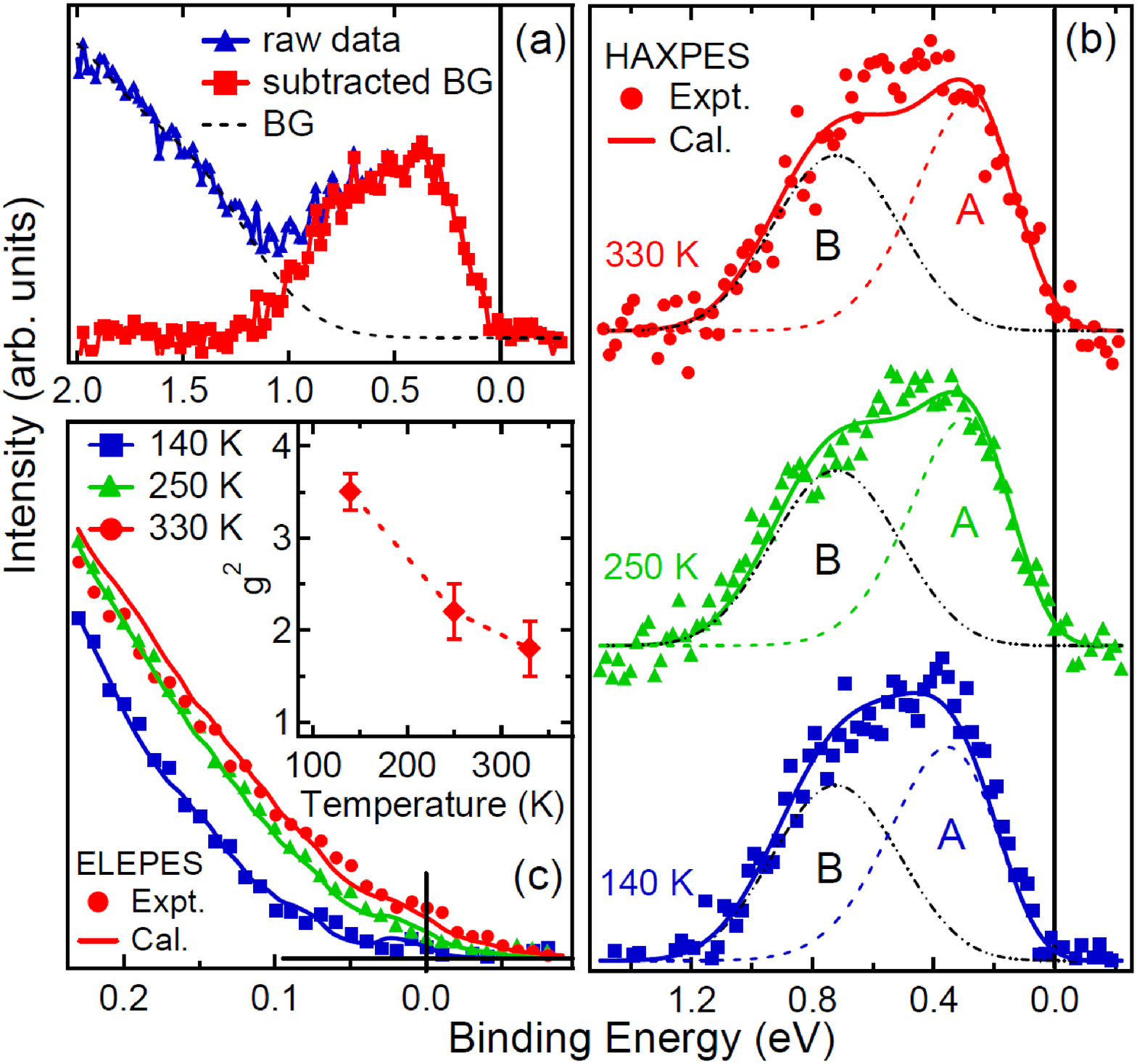}
\end{center}
\caption{(color online) (a) Raw HAXPES spectrum near $E_F$  (blue curve) and the background-subtracted spectrum (red curve) of Fe$_3$O$_4$ at 140 K (above $T_V$). (b) Model fit of the background-subtracted HAXPES spectra (see text for the label A and B). (c) Model fit of the ELEPES spectra with the same parameters as HAXPES spectra. Inset shows the temperature dependence of the dimensionless electron-phonon coupling constant $g^2$. }
\label{f3}
\end{figure}

We have tried to analyze the HAXPES and ELEPES results by modeling the above polaronic spectral function based on eq. (1) and the LDA calculation.\cite{lda3} The LDA DOS near $E_F$ consists of two components, one crossing $E_F$ composed of minority spin states and the other one centered at $E_B$ = 0.6 eV composed of majority spin states. The polaronic model is applied only to the former component, because the spectral change around $E_B$ = 0.6 eV is much smaller than that near $E_F$. In order to reproduce the HAXPES (ELEPES) spectra in Fe$_3$O$_4$, we have taken the photoionization cross section ratio between Fe 3d and O 2p states into account\cite{cross_section_ratio} and broadened the theoretical spectrum by 120 meV (10 meV) to consider the instrumental resolution. We have first subtracted the Fe 4s + O 2p contribution above $E_B$ = 1 eV in HAXPES as shown in Fig. 3(a) by the dashed curve and then approximated the component centered at $E_B$ = 0.6 eV by a proper Gaussian peak.\cite{gaussian} According to the results of the infrared study for Fe$_3$O$_4$ two phonon modes were reported,\cite{OC1,OC2} one at $\sim $40 meV and the other at $\sim $70 meV. We have tried to fit the spectra with either phonon energy because it is not known which phonon mode couples to the electron. A good agreement between the calculation and the experiments is obtained by using the higher phonon energy of 70 meV as in the case of SXPES results. Therefore, the characteristic phonon energy $\omega _0$ is fixed to 70 meV. The coupling constant $g^2$ is treated as the fitting parameter. We note that the temperature dependence of the HAXPES and ELEPES spectra in HT phase could not be satisfactory explained without changing $g^2$ with temperature as discussed below.

Figure 3 (b) and (c) show the fitting results of the background-subtracted HAXPES spectra and the ELEPES spectra. In Fig. 3 (b), the dashed curves (Peak A) show the spectra derived through eq. (1) from the DOS component crossing $E_F$ and the dot-dashed curves (Peak B) show the spectra from the DOS centered at $E_B$ = 0.6 eV. We can well reproduce the spectra near $E_F$ at each temperature with the common coupling strength $g^2$ of 3.4, 2.2 and 1.8 at 140, 250 and 330 K through HAXPES and ELEPES. Such a temperature-dependent coupling strength has been implied by the optical conductivity.\cite{OC1,OC2} The coupling strength $g^2$ is plotted as a function of temperature in the inset of Fig. 3 (c). At 140 K, $e^{-3.4}$ $\sim $ 0.033 means the suppression of the spectral weight at $E_F$ of more than 96 $\%$, where the value of the $g^2$ = 3.4 [$\varepsilon _p$ = 238 meV being consistent with the polaronic binding energy ($\sim $300 meV) derived from the infrared spectroscopy\cite{OC1,OC2}] reflects the almost insulating feature and the formation of the small polaron.  Then $g^2$ ($\varepsilon _p$) gradually decreases with increasing temperature. The suppression of the spectral weight due to the formation of the polaron becomes reduced as $e^{-2.2}$ $\sim $ 0.11 and $e^{-1.8}$ $\sim $ 0.17 around 250--330 K, indicating the variation from the small polaron to the large polaron with increasing temperature. It has been suggested from the results of neutron scattering,\cite{neutron_scattering} $\mu $SR,\cite{muSR} transport phenomena,\cite{transport} and M\"{o}ssbauer spectroscopy\cite{Mossbauer} that the change of the electronic states takes place below around 250-300 K  corresponding to the formation of the large polaron. Our results strongly support the scenario as the variation takes place from the small polaron to the large polaron with increasing temperature in the HT phase of Fe$_3$O$_4$.

In conclusion, we have performed HAXPES and ELEPES for Fe$_3$O$_4$ and observed the detailed spectral changes with increasing the temperature. In the HT phase just above the Verwey transition, the intensity at $E_F$ is still negligible. However, it increases gradually with increasing temperature. This temperature dependence can be well explained by considering the polaron effect. Since the polaronic coupling strength becomes weaker at high temperatures far from $T_V$, the intensity near $E_F$ increases inducing the increased conductivity of Fe$_3$O$_4$ at higher temperatures.

This work was supported by the Grant-in-Aid for Science Research (18104007, 18684015, 21340101, 21740229, "heavy electron" 20102003) of MEXT, Japan, and the Global COE program (G10) of JSPS, Japan. A part of the ELEPES study was performed as a Joint Studies Program of the Institute for Molecular Science (2007).


\begin{thebibliography}{99}
\bibitem{cross_section_ratio} The cross-section ratio between Fe 3d and O 2p has been set to 2 : 1 (1 : 11) for HAXPES (ELEPES). The ratio for HAXPES is based on the values in Ref. 29. We have estimated the photoionization cross sections at $h\nu $ = 7.5 eV by extrapolating the values at $h\nu $ $>$ 10.2 eV\cite{cross_section} since there is no calculation of the cross sections at $h\nu $ = 7.5 eV.
\bibitem{gaussian} Here we have thought that the shape and position of the polaronic behavior is important near $E_F$, because they are dominated by the polaron effect in the component crossing $E_F$. Therefore, a simple Gaussian fitting for the component at $E_B$ = 0.6 eV away from $E_F$ is fully sufficient.
\end{thebibliography}
\end{document}